\documentclass[11pt, aps,prl, superscriptaddress,amsmath,longbibliography,amssymb,preprint]{revtex4-1}
\usepackage{graphicx} % Include figure files
\usepackage{dcolumn}  % Align table columns on decimal point
\usepackage{bm}       % bold math
\usepackage{amsfonts}
\usepackage{color}

\usepackage{soul,xcolor,easyReview}
\usepackage{ulem}
\usepackage{color}

\begin{document}
\preprint{}
\title{Nonlinear Optical Spectroscopy of Nodal-Line Semimetals}

\author{Navdeep Rana}
\email{navdeeprana@lsu.edu}
\affiliation{%
            Department of Physics and Astronomy, Louisiana State University, Baton Rouge, Louisiana 70803-4001, USA}
\author{M. S. Mrudul}
\affiliation{%
            Department of Physics and Astronomy, Uppsala University, P.O. Box 516, SE 75120 Uppsala, Sweden}
\author{Amar Bharti}
\affiliation{%
            Department of Physics, Indian Institute of Technology Bombay, Powai, Mumbai 400076 India}
\author{Sucharita Giri}
\affiliation{%
            Department of Physics and Astronomy, Louisiana State University, Baton Rouge, Louisiana 70803-4001, USA}
\author{Gopal Dixit}
\email{gdixit@phy.iitb.ac.in}
\affiliation{%
            Department of Physics, Indian Institute of Technology Bombay, Powai, Mumbai 400076  India}
\affiliation{%
	    Max-Born Institute, Max-Born Stra{\ss}e 2A, 12489 Berlin, Germany.}
\date{\today}
%\pacs{}

%%%%%%%%%%%%%%%%% END OF PREAMBLE %%%%%%%%%%%%%%%%

\begin{abstract}
Intense laser-driven nonlinear optical phenomena in two-dimensional (2D) nodal-line semimetals (NLS) exhibit complex mechanisms, particularly in the NbSi$_{x}$Te$_{2}$ material systems characterized by nonsymmorphic symmetry-protected band degeneracy. 
Our findings reveal how nonsymmorphic symmetry-protected band degeneracy fundamentally influences the material's nonlienar optical responses. 
Notably, the nonsymmorphic glide-mirror symmetry leads to the exclusive generation of odd-order harmonics from inversion-symmetry-broken NLS. 
Moreover,  harmonics are emitted parallel and perpendicular to the driving laser's polarization. 
We demonstrate distinct generation mechanisms arise from intrachain and interchain processes, with their relative contributions varying significantly with the polarization of the driving laser pulse. 
The polarization-dependence exhibits  two-fold anisotropy, with each harmonic order showing characteristic angular distributions of maximum yield. 
Additionally, our analysis of the ellipticity-dependence reveals an intricate interplay between interband and intraband mechanisms. 
These insights open new possibilities for controlling harmonic generation through precise tuning 
parameters of the driving laser and highlight the potentials of NLS materials to  
%chiral-sensitive light-matter interactions,  
fabricate lightwave-based photonics, optoelectronic and quantum devices operating on ultrafast timescales. 	
\end{abstract}

\maketitle

%\section{Introduction} 
 The groundbreaking discovery of monolayer graphene has stimulated extensive research activities in recent years~\cite{novoselov2004electric}. 
One of the distinctive properties of graphene is linear band dispersion along high-symmetry points at which the conduction and valence bands touch each other. 
The presence of the band-touching linear dispersion in two-dimensional (2D) graphene has motivated researchers to synthesize analogous materials in three dimensions~\cite{armitage2018weyl, hasan2010colloquium}.  
Dirac and Weyl semimetals are three-dimensional equivalents of graphene in which low-energy dispersions of the conduction and valence bands meet at isolated points in momentum space~\cite{xu2015discovery, xu2015discovery1, lv2015experimental}. 
The nodal points in Dirac and Weyl semimetals are topological in nature, and the points
need not be along high-symmetry points. 
The presence of isolated nodal points can be translated into a line along which
both valence and conduction bands touch. 
The appearance of the band-touching dispersive line in momentum space
is caused by the presence of nonsymmorphic symmetry in certain materials~\cite{li2018nonsymmorphic, sato2018observation, chen2015nanostructured, burkov2011topological}. 
Materials with  band-touching  dispersive nodal line(s) in energy dispersion are known as nodal-line semimetals (NLS)~\cite{chiu2016classification, fang2016topological, gao2018epitaxial, shih2024quantized, barati2017optical}. 
Quantum semimetals with nodal points and nodal lines are essential  
for emerging conceptions of next-generation electronic, optoelectronic and quantum devices~\cite{sirica2021shaking, tokura2017emergent, mrudul2021light, keimer2017physics,   burkov2016topological}. 
Techniques relying on nonlinear laser-matter interaction have proven critical  to unravel various exotic properties of  quantum semimetals~\cite{basov2017towards, bao2021light, bharti2023weyl, mciver2012control, yan2017topological, bharti2024photocurrent, rana2024optical}. 
In this respect,  how the presence of the nonsymmorphic symmetry  and resultant nodal line affect  
different optical properties of NLS are least explored territory~\cite{sinha2021giant, tavakol2023nonlinear}. 
The present work addresses such crucial questions by investigating nonlinear optical properties of 
nodal-line semimetals.

High-harmonic generation (HHG) is a nonperturbative nonlinear optical process, which allows  generation of  coherent radiation in extreme-ultraviolet and soft x-ray regimes. 
Over the decades, HHG has become a popular method to produce attosecond pulses and interrogate electron dynamics in matter~\cite{krausz2009attosecond}. 
Owing to the immense importance of it, the 2023 Nobel Prize in Physics  was awarded for HHG and 
attosecond physics~\cite{krausz2024nobel, agostini2024nobel, l2024nobel}.
The pioneering work of Reis and co-workers has successfully extended HHG from gases to solids~\cite{ghimire2011observation}. 
Since its inception, solid HHG is a preferable method to probe various nonequilibrium aspects of the interrogated solids~\cite{luu2015extreme,   heide2022probing, langer2018lightwave, pattanayak2019direct, imai2020high, borsch2020super, vampa2015all,  schmid2021tunable, qian2022role,  silva2018high,  yue2020imperfect, ghimire2019, goulielmakis2022high, mrudul2020high, pattanayak2020influence}.  
In addition, HHG from quantum semimetals with nodal points has spurred  enormous research interest  in recent years~\cite{yoshikawa2017high, taucer2017nonperturbative, cha2022gate,  rana2022probing, avetissian2022efficient, murakami2022doping, dong2021ellipticity, zhang2021orientation, boyero2022non, guan2023optimal, rana2022high, dong2022knee, zurron2019optical, mrudul2024dependence, zurron2018theory, liu2018driving, sorngaard2013high, ishikawa2010nonlinear, chizhova2016nonlinear, mikhailov2007non, avetissian2012creation, rana2022generation, avetissian2018impact, li2022high, bharti2024non, avetissian2022high, lv2021high, wang2024table,  kovalev2020non, cheng2020efficient, lim2020efficient}.

\begin{figure}
\includegraphics[width= 0.8 \linewidth]{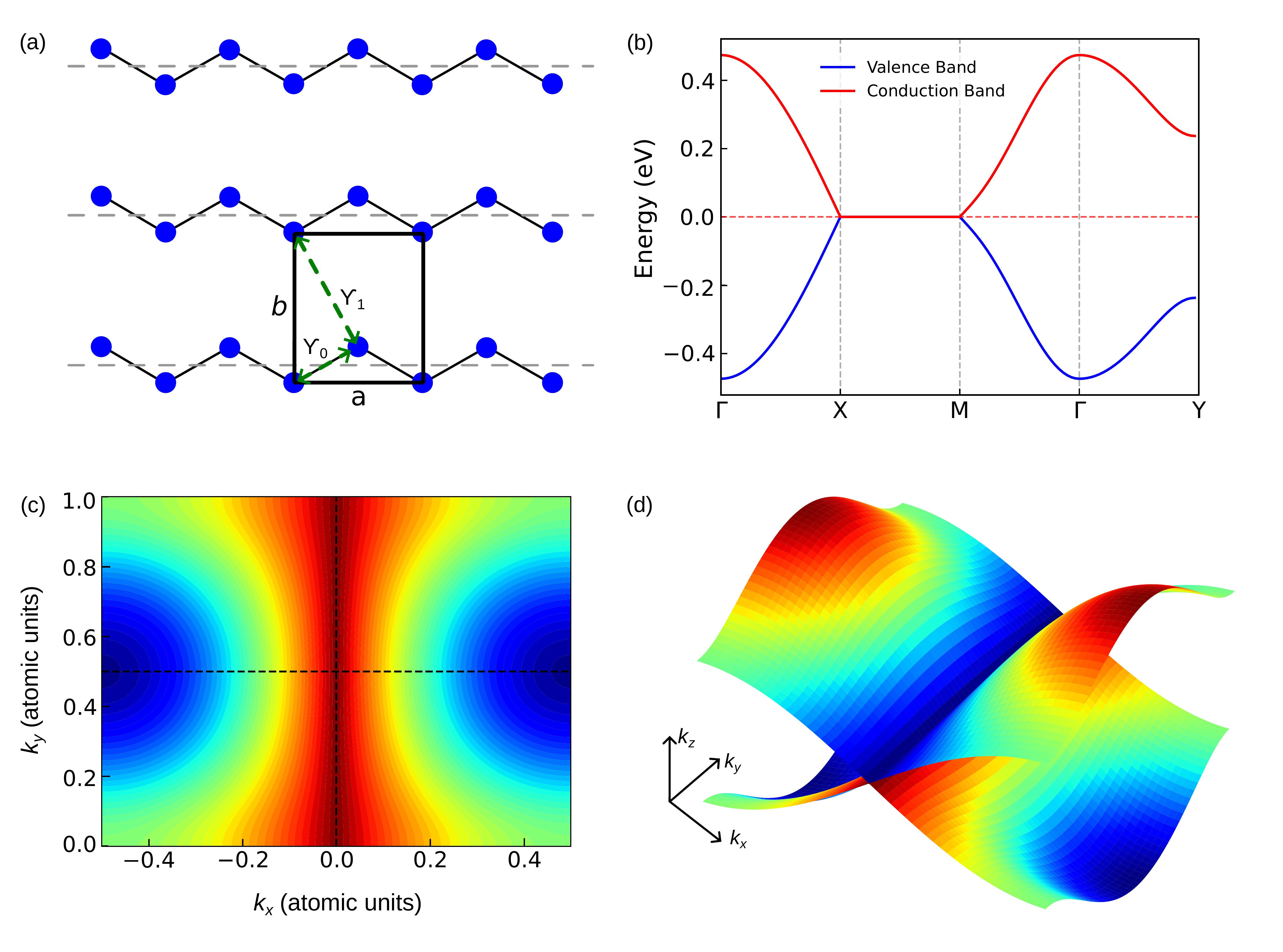}
\caption{(a) Real-space lattice structure of the nodal-line semimetal in 2D.  
The black rectangle highlights  the unit cell, where $\gamma_{0}$ and $\gamma_{1}$ represent intrachain (within a chain) and interchain (between the chains) hopping. 
(b) Energy band dispersion of the  nodal-line semimetal along high-symmetry directions.   
(c) Two-dimensional  and (d)  three-dimensional views of the energy dispersion.}
~\label{schematic}
\end{figure}

In this Letter, we illustrate key signatures of the nonsymmorphic symmetry in high-harmonic spectroscopy. 
For this purpose, we consider 2D NLS in which nodal lines are along high-symmetry directions in the Brillouin zone with strong anisotropy. 
Recently,  the family of NbSi$_x$Te$_2$ materials with $x$ = 0.40 and 0.43 has been found to exhibit features  of  2D NLS~\cite{zhang2022observation, wang2021one, zhu2020tunable, yang2020directional}. 
In these materials, NbTe$_2$ forms zigzag chains with two lattice sites, A and B, which are related by a nonsymmorphic symmetry.  
The coupling between different zigzag chains leads to nodal lines in the energy band structure [see Figs.~\ref{schematic}(a) and (b)]. 
In the following, we demonstrate that linearly polarized light  along $x$- and $y$-directions  results in the   
generation of elliptically polarized harmonics in 2D NLS.
Emission of harmonics with varying ellipticity can be traced to intriguing interplay between interband and intraband dynamics. 
In addition, polarization dependence of the emitted harmonics exhibits strong twofold anisotropy. 
These observations for 2D NLS are drastically different from the known results of monolayer graphene with nodal points.  

%section{Theoretical Method}
We consider a zigzag chain along the $x$ direction with a lattice constant $a$, 
and the lattice period along the $y$ direction is represented by $b$, which is the interchain distance as evident  in Fig.~\ref{schematic}(a).
The electronic structure of the 2D NLS within the tight-binding description  can be expressed  as ~\cite{zhang2022observation, cao2023plasmons}
\begin{equation}\label{eqtb}
\mathcal{H}(\mathbf{k}) = 
\gamma_{0}  \left(~ 1+ e^{- ik_{x}a}~\right)~\hat{\alpha}_\mathbf{k}^{\dagger} \hat{\beta}_\textbf{k}+ 	 \gamma_{1}e^{-ik_{y}b} \left(~ 1+ e^{-ik_{x}a}~\right)~\hat{\alpha}_\textbf{k}^{\dagger} \hat{\beta}_\mathbf{k} + \textrm{H. c.}
\end{equation}
Here, the first term corresponds to the intrachain coupling, whereas the second term stands for the interchain coupling as shown in Fig.~\ref{schematic}(a). 
$\gamma_{0} (\gamma_{1})$ = 0.18 (0.05) eV is the intrachain (interchain) coupling constant,  
$\hat{\alpha}_k^{\dagger}~(\hat{\beta}_k)$  
represents the creation (annihilation) operator for atom A (B) in the unit cell
 and $\textrm{H.c.}$ stands for Hermitian conjugate.  
The two sites A and B within the unit cell are related by the glide mirror symmetry ($\tilde{M_{y}}$) as visible from  Fig.~\ref{schematic}(a).
The energy dispersion is obtained by diagonalization of the Hamiltonian  as 
\begin{equation}
\mathcal{E}(\mathbf{k}) = \pm 2 \cos\left({\dfrac{k_{x}a}{2}}\right) \sqrt{\gamma_{0}^{2} + \gamma_{1}^{2} + 2\gamma_{0}\gamma_{1}\cos({k_{y}b})}.
\end{equation}
Here, $\pm$ represents the conduction and valence bands of the 2D NLS.  
The energy band structure along the high-symmetry direction exhibits degenerate bands along the \textsf{X-M} path as reflected from Fig.~\ref{schematic}(b). 
Note that the appearance of the nodal line is robust against the strength of the interchain coupling as the glide mirror symmetry ($\tilde{M_{y}}$) is intact. 
%The Hamiltonian in Eq.~(\ref{eqtb}) preserves both time-reversal and inversion symmetries. 

The density-matrix-based equation of motion within the Houston basis is employed to describe the interaction of an intense laser pulse with NLS~\cite{wilhelm2021semiconductor, rana2023all, yue2022introduction, bharti2023tailor}.
The total electronic current $\mathbf{J}(t) $ in the Brillouin zone is simulated by solving coupled Bloch equations in momentum space as discussed in the Supplemental Material~\cite{NoteX}. 
The high-harmonic spectrum is obtained by performing the Fourier transform of the time derivative of total current integrated over the entire Brillouin zone as 
\begin{equation}
\mathcal{I}(\omega) = \left| \mathcal{FT} \left(  \dfrac{d}{dt} \mathbf{J}(t)    \right) \right|^2.
\end{equation}
Here, $\mathcal{FT}$ stands for the Fourier transform.  
In this work,  high-order harmonics are generated using a laser pulse with a wavelength of 4.8 $\mu$m and peak intensity of $10^{10} $ W/cm$^2$. 
The pulse is 85 fs long and has a sine-squared envelope.  
A constant phenomenological parameter of 10 fs is used to account for the decoherence between electrons and holes. 
Our results are qualitatively the same for decoherence parameter ranging from 1 to 20 fs (see Fig. S1 in the Supplemental Material~\cite{NoteX}). 
These laser parameters have been used previously to generate harmonics in graphene~\cite{mrudul2021high,bharti2023role,rana2024high}.

%\section{Results and Discussion}
\begin{figure}
\includegraphics[width= 0.9\linewidth]{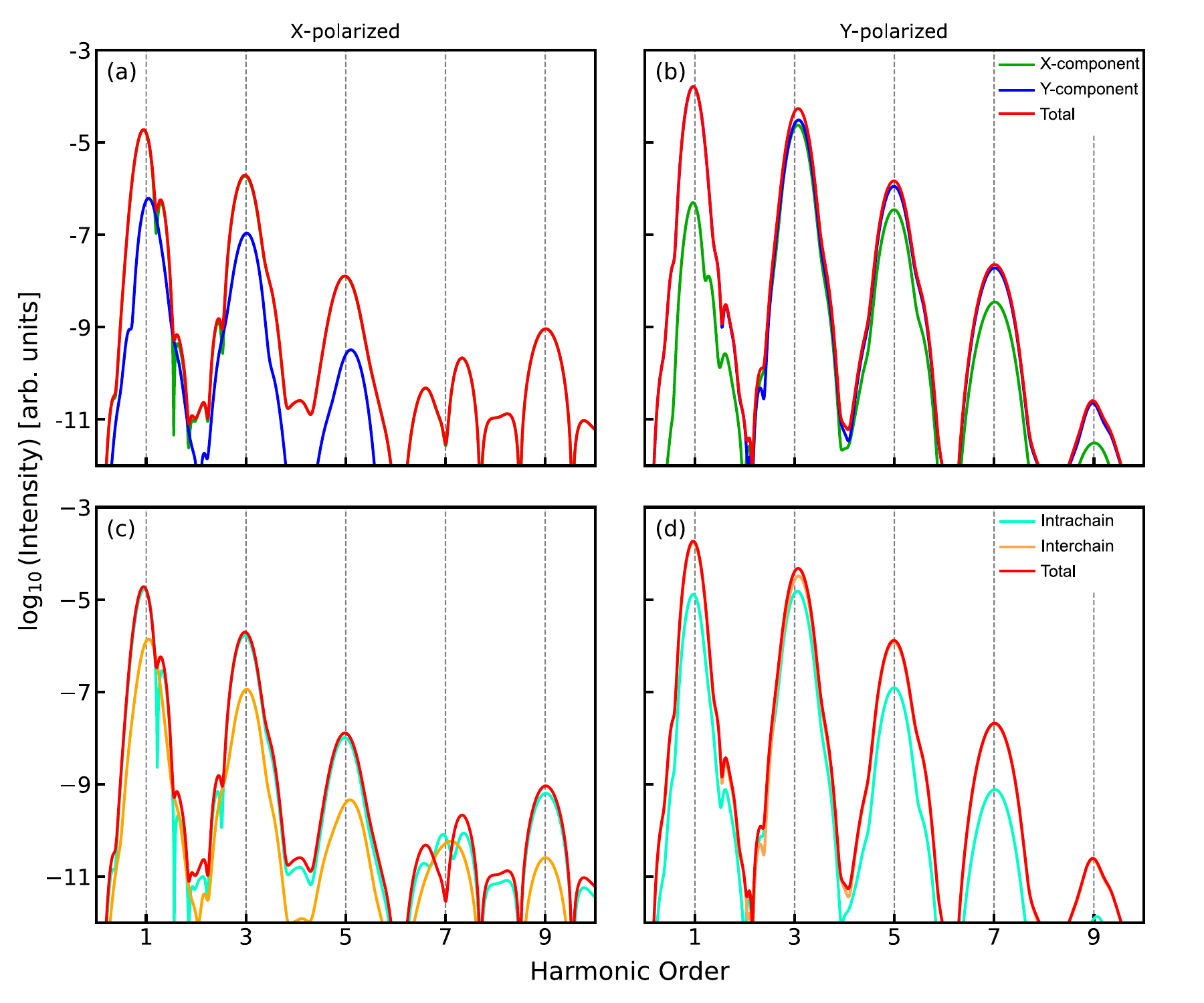}
\caption{High-harmonic spectra of a two-dimensional nodal-line semimetal driven by a linearly polarized laser pulse along the (a), (c) $x$- and (b), (d) $y$-directions. Panels (a) and (b) show direction-resolved spectra, while panels (c) and (d) present intrachain and interchain contributions. The red line represents the total harmonic intensity, with green and blue lines representing  emitted harmonics polarized along the $x$- and $y$-directions in  (a) and (b), respectively. The intrachain and interchain contributions  in (c) and (d) are depicted by cyan and orange  lines, respectively.}
\label{spectrum}
\end{figure}

Let us start by analyzing the harmonic spectra for laser polarization along the $x$-direction as shown in 
Fig.~\ref{spectrum}(a).  
As evident, odd-order 
harmonics are generated  both parallel and perpendicular to the driving laser polarization. 
While the parallel and perpendicular components have comparable strength  
for the lower-order harmonics, the parallel harmonics begin to dominate as the harmonic order increases.
The total harmonic spectrum changes significantly as 
the laser polarization switches to the $y$-direction as reflected  in Fig.~\ref{spectrum}(b). 
However, the general observations remain the same, i.e.,  
harmonics are generated  both parallel and perpendicular to the laser polarization  irrespective of the laser's polarization.
It is known that both odd- and even-order harmonics are  generated for inversion-broken materials~\cite{liu2017high}. 
In addition, inversion-broken systems result in Berry-curvature-driven 
even-order harmonics, which are  polarized perpendicular to the driving laser's polarization~\cite{luu2018measurement}.
However, it is not obvious {\it a priori }  why only the odd-order harmonics are generated in an inversion-broken 2D NLS as reflected in Figs.~\ref{schematic}(a) and ~\ref{spectrum}.

To address the underlying physical mechanism responsible for the selective generation of only odd-order harmonics in inversion-broken 2D NLS, let us analyze symmetry properties of  $\mathcal{H}(\textbf{k})$. The Hamiltonian 
exhibits nonsymmorphic glide-mirror symmetry ($\tilde{M_{y}}$) as 
the key lattice symmetry.  
The $\tilde{M_{y}}$ symmetry operation essentially exchanges atoms of two sublattices (A$\leftrightarrow$B), similar to the inversion symmetry operation in graphene. Thus, the  $\tilde{M_{y}}$ symmetry acts as a pseudo-inversion-symmetry in $\mathcal{H}(\textbf{k})$. 
Mathematically, the transformation of $\mathcal{H}(\textbf{k})$ corresponding to $\tilde{M_{y}}$ is written as $\mathcal{H}(\textbf{k}) = \sigma_{x} \mathcal{H}(-\textbf{k}) \sigma_{x}$ with $\sigma_{x}$  the Pauli matrix.
In addition to spatial symmetries, the system also exhibits a dynamical symmetry associated with the drive: $\mathbf{A}(t + \textrm{T}/2) = -\mathbf{A}(t)$. 
This constrains the time-dependent Hamiltonian as $\mathcal{H}(\textbf{k}_{t}) = \sigma_x \mathcal{H}(-\textbf{k}_{t+\textrm{T}/2}) \sigma_x$.
Using the dynamical symmetries, as outlined in the Supplemental Material~\cite{NoteX}, we can demonstrate 
\begin{equation}
\mathbf{J}_{a}^{i}(\textbf{k},n\omega) = \int_{0}^{\textrm{T}} \dfrac{dt}{\textrm{T}}~e^{in\omega}~\mathbf{J}_{a}^{i}(\textbf{k},t+\textrm{T}/2) = -~e^{in\pi}~\mathbf{J}_{a}^{i}(\textbf{k},n\omega),
 \label{Srule}
\end{equation}
where $\mathbf{J}_{a}^{i}(\textbf{k}, n\omega)$ represents the $n^{\textrm{th}}$ harmonic amplitude along the $i^{\textrm{th}}$ direction,  $i$ can be $x$ or $y$, $\omega$ is the frequency, and ${\textrm{T}}$
is the  time period of the driving laser. 
It is straightforward to deduce that $\mathbf{J}_{a}^{i}(\textbf{k}, n\omega) = 0$ for even values of $n$ in both the $x$- and $y$-directions. Therefore, the exclusive observation of the odd-order harmonics serves as the unique signature of the  nonsymmorphic glide-mirror symmetry in 2D NLS.

After establishing key signatures of the nonsymmorphic symmetry in HHG, let us examine contributions from intrachain and interchain processes to unravel the underlying mechanism responsible for generating perpendicular harmonics. 
The current operator, defined as $\hat{\textbf{J}}(\textbf{k}) = \nabla_\mathbf{k} \mathcal{H}(\mathbf{k})$, can be seperated into interchain and intrachain parts, as done for the Hamiltonian in Eq.~(\ref{eqtb}).
Figures~\ref{spectrum}(c) and~\ref{spectrum}(d) illustrate the intrachain- and interchain-resolved 
high-harmonic spectra for linearly polarized pulses along the $x$- and $y$-directions, respectively. 
It is evident that the relative contributions stemming from the  intrachain and interchain to total harmonics are intricate. 
The spectrum corresponding to the $x$-polarization is dominated by the intrachain contributions 
and the interchain contributions diminish  significantly for higher orders as shown in Fig.~\ref{spectrum}(c). 
However, the interchain contributions dominate for  the $y$-polarized laser and the 
intrachain contributions decrease as harmonic orders increase [see Fig.~\ref{spectrum}(d)].  
Interchain contribution   
for the $x$-polarized laser results in
harmonics  parallel and perpendicular to the laser polarization, which 
exhibit characteristic  polarization of the emitted harmonics. 
Additionally, the interchain contribution diminishes with increasing order, 
resulting in higher-order harmonics being polarized exclusively parallel to the laser pulse. 
A similar reasoning for the $y$-polarized laser is applicable for which  intrachain harmonics lead to
additional perpendicular components in the spectra.

\begin{figure}
\includegraphics[width= 0.5\linewidth]{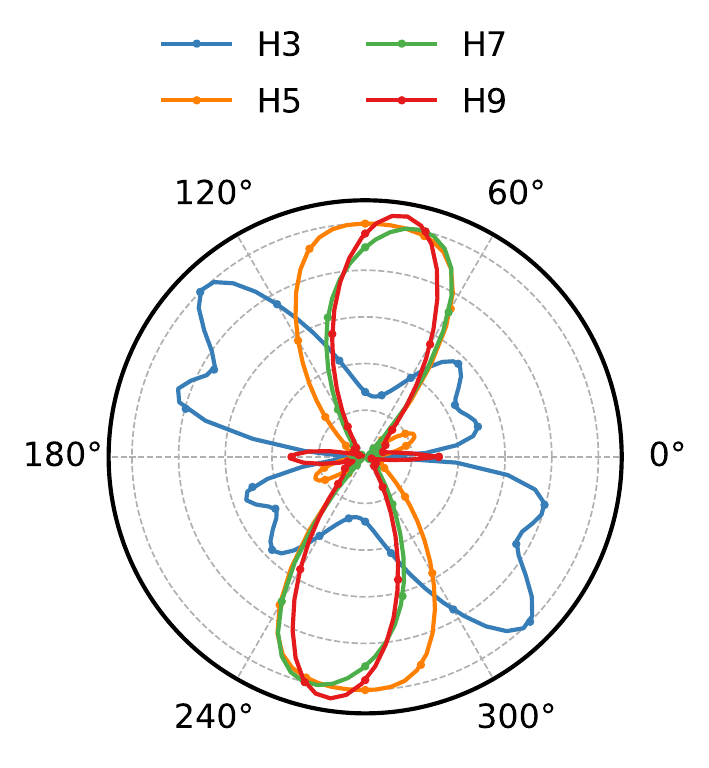}
\caption{Polarization dependence of the normalized harmonic yield for the third (H3),  fifth (H5), seventh (H7) and ninth (H9) harmonics. 
The blue, orange, green and red colors represent the normalized yields of
H3, H5, H7 and H9, respectively.}
\label{pol_depend}
\end{figure}

Building upon our analysis stemming from the linear pulse along $x$- and $y$-directions, it is imperative to know how the presence of the nodal line affects  the polarization dependence of HHG.  
The vector potential of a linearly polarized pulse is written as 
\begin{equation}
\mathbf{A}(t) =  A_{0} f(t) \cos({\omega t})  \left[  \cos({\theta})~\hat{\mathbf{e}}_{x} + \sin({\theta})~ \hat{\mathbf{e}}_{y}  \right],
\end{equation}
where $A_{0}$ is the amplitude and $f(t)$ is the envelope of the vector 
potential, and $\theta$ stands for the angle between the $x$ axis and laser polarization. 
Figure~\ref{pol_depend} shows  anisotropic  polarization dependence of the 
normalized harmonic yield for the third (H3),  fifth (H5), seventh (H7) and ninth (H9) harmonics. 
Notably, the polarization dependence displays a twofold symmetry in the total yield.
In addition, the harmonic yield exhibits maxima around 135$^\circ$ for H3, and 90$^\circ$ for H5, 75$^\circ$ 
for H7, and 80$^\circ$ for H9, indicating a preferential generation of the maximum harmonic's yield  at these specific polarization angles.
The present results are for laser's intensity 10$^{11}$ W/cm$^2$
and remain consistent at 10$^{10}$ W/cm$^2$ with the same twofold symmetry and
preferential direction, though the maximum angle may varies.
Additionally, the present observations are in contrast to those for graphene in which H3 is isotropic in nature, whereas H5 and H7 display sixfold symmetry -- mimicking the symmetry of graphene with nodal points~\cite{mrudul2021high}.   
Thus, the lattice structure of the 2D 
NLS and corresponding anisotropic nodal lines can be attributed as the key reason for the twofold anisotropic polarization dependence~\cite{kandel2023anisotropic}. 

\begin{figure}
\includegraphics[width=  0.9 \linewidth]{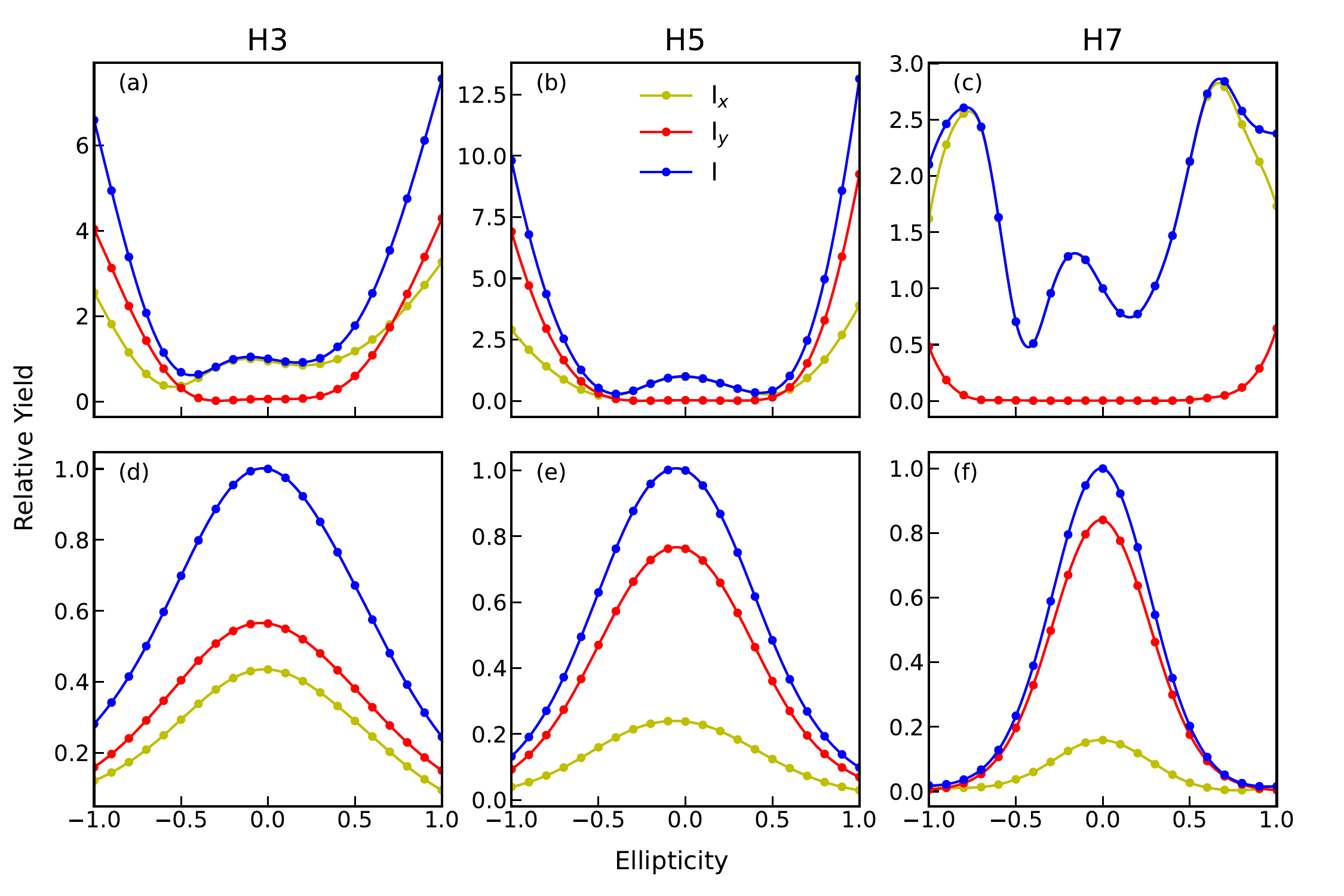}
\caption{Integrated harmonic yield as a function of the incident laser's ellipticity for the (a) third (H3), (b) fifth (H5) and (c) seventh (H7) harmonics. The major axis of the laser pulse is along the $x$-direction. 
(d), (e) and (f) The same as (a), (b) and (c) for the driving pulse with the $y$-direction as the major axis. 
Red, blue, and green lines represent the normalized integrated yield for the $x$-, $y$-, and total harmonic components, respectively. 
The total harmonic yield at  finite ellipticity is normalized with respect to the harmonic yield for zero ellipticity. } \label{ellipticity}
\end{figure}

With the critical role of the polarization dependence revealed, the next intriguing aspect is to explore  ellipticity dependence of HHG. 
Let us investigate  how harmonics' yield  is responsive  to the ellipticity of the driving  laser. 
The vector potential of  the  driving laser pulse with ellipticity ($\epsilon$)  is written as 
\begin{equation}
\mathbf{A}(t) =  \dfrac{A_{0} f(t)}{\sqrt{1+\epsilon^{2}}}  \left[ \cos({\omega t}) ~\hat{\mathbf{e}}_{x} + \epsilon \sin({\omega t})~ \hat{\mathbf{e}}_{y}  \right].
\end{equation}
 Here, $\epsilon$ of value 1 implies a right-handed circularly polarized laser  and -1 implies a left-handed circularly polarized laser.

The integrated yields of H3, H5, and H7 as a function of $\epsilon$ are shown in Figs.~\ref{ellipticity}(a) - (c), respectively. 
In this case, the driving laser has its major axis aligned along the $x$-direction. 
Not only does the total harmonic yield exhibit a distinctive response to the laser's ellipticity, 
but the component-resolved yields along $x$- and $y$-directions are also sensitive to the ellipticity.
Notably,  H3 and H5 display a distinct ellipticity dependence compared with H7. 
It is evident that both H3 and H5, along with their respective components, exhibit maximum yield for a circularly polarized pulse. 
H7 demonstrates completely  different behavior: the total yield and its $x$-component have maximum for ellipticity value of $\pm 0.85$ and minimum around ellipticity value of $ - 0.45$, beyond which it starts  to increases again. 
The $y$-component of H7, however, remains nearly insensitive to ellipticity up to $\pm 0.8$, after which it begins to exhibit a strong dependence as ellipticity increases further as evident from Fig.~\ref{ellipticity}(c).

The ellipticity-dependence changes significantly as the major axis of the ellipse  changes from the $x$- to the $y$-direction.  
The harmonics H3, H5 and H7 display similar ellipticity dependence as evident in Figs.~\ref{ellipticity}(d) - (f), respectively. 
Their yields, along with their $x$- and $y$-components, are maximized for linearly polarized pulses and decrease monotonically with increasing $\epsilon$, approaching minimum values for H3 and H5,  and zero for H7 under circular polarization.  
The contrasting behavior of H3, H5, and H7 when driven by laser pulses with major axes along the $x$- and $y$-directions can be attributed to the interplay between different hopping terms and the lattice structure of the 2D NLS. 
This demonstrates that the ellipticity dependence is highly sensitive to the nodal line structure and its inherent nonsymmorphic symmetry.

The characteristic features in the harmonic yields  are inherently related to the underlying mechanisms responsible for the harmonic generation. 
The lower-order harmonics have contributions from both the interband and intraband harmonics, whereas the higher-order harmonics are dominated by the intraband component in 2D NLS (see Fig. S2 in the Supplemental Material~\cite{NoteX}).
The characteristic ellipticity dependence arises because interband transitions between conduction and valence  bands are independent of the electric field's steering in momentum space. 
However, the intraband mechanism is associated with the  group velocity and directly probes the conduction band dispersion through the driving laser field. 
The interplay between two mechanisms leads to varied yields for each harmonic and the observed diversity in ellipticity profiles. 
Analysis of Fig.~\ref{ellipticity} establishes the idea of  tailoring the harmonic yield at a finite laser ellipticity, potentially due to an efficient coupling between the driving laser and the solid~\cite{you2017anisotropic}. 
Our findings are in line with the notion that interband and intraband processes exhibit varying degrees of sensitivity to the laser's ellipticity as discussed in Refs.~\cite{tancogne2017ellipticity,you2017anisotropic}.

%\section{Conclusion} 
 In summary, the harmonic spectra in 2D nodal-line semimetals provide  key insights into their inherent nonsymmorphic symmetry and lattice structures.
The harmonic spectra show distinct contributions from intrachain and interchain processes, with intrachain contributions dominating for laser polarization along the $x$-direction, while interchain contributions are more prominent for $y$ polarization. 
The exclusive generation of the odd-order harmonics is attributed to the nonsymmorphic 
glide-mirror symmetry, which forbids even-order harmonics despite the absence of an inversion symmetry. 
Our study further reveals the intricate polarization and ellipticity dependences of the harmonic yield. 
Polarization dependence of the different harmonics' yield displays a  twofold anisotropy, with preferential angles of the maximum yield for different harmonic orders. 
Additionally, the yields' sensitivity to the ellipticity of the driving laser highlights the influence of the nonsymmorphic symmetry of the material on the harmonic generation, with distinct responses for the third, fifth, and seventh harmonics. 
These findings illuminate the complex interplay between interband and intraband mechanisms, demonstrating that the lower-order harmonics are less sensitive to the laser's steering in momentum space, whereas higher-order harmonics 
are directly influenced by the conduction band dispersion and the group velocity of electrons. 
The observed variations in ellipticity and polarization profiles suggest potential avenues for controlling and optimizing harmonic generation by tailoring driving laser parameters. 
Our work highlights the crucial role of a material's structural and  symmetry properties in shaping harmonic spectra, providing insights to  advance  new avenues for various nonlinear optical processes in quantum semimetals.
Our model, based on a Dirac-SSH Hamiltonian for NbSi$_x$Te$_{2}$ ~\cite{zhang2022observation}, captures the nonsymmorphic symmetry and is consistent with {\it ab initio} results, motivating future exploration of multiband and topological effects in the harmonic response.

%\section{Acknowledgements}

G.D. acknowledges financial support from SERB India (Project No. MTR/2021/000138).
%The data that support the findings of this article are openly available ~\cite{rana_2025_15653873}.
%\bibliography{solid_HHG}


\end{document}